# BlurryScope enables compact, cost-effective scanning microscopy for HER2 scoring using deep learning on blurry images

***Michael John Fanous[1], Christopher Michael Seybold[2,#], Hanlong Chen[1,3,4,#], Nir Pillar[1,3] and Aydogan Ozcan[1,3,4,5,*]***

*[1]Electrical and Computer Engineering Department, University of California, Los Angeles 90095 CA, USA*
*[2]Mathematics Department, University of California, Los Angeles 90095 CA, USA*
*[3]Bioengineering Department, University of California, Los Angeles 90095 CA, USA*
*[4]California NanoSystems Institute (CNSI), University of California, Los Angeles 90095 CA, USA*
*[5]Department of Surgery, David Geffen School of Medicine, University of California, Los Angeles 90095 CA, USA*

* *ozcan@ucla.edu*

*[#]C.M.S and H.C: these authors contributed equally to this work.*

## Abstract

We developed a rapid scanning optical microscope, termed "BlurryScope", that leverages continuous image acquisition and deep learning to provide a cost-effective and compact solution for automated inspection and analysis of tissue sections. This device offers comparable speed to commercial digital pathology scanners, but at a significantly lower price point and smaller size/weight. Using BlurryScope, we implemented automated classification of human epidermal growth factor receptor 2 (HER2) scores on motion-blurred images of immunohistochemically (IHC) stained breast tissue sections, achieving concordant results with those obtained from a high-end digital scanning microscope. Using a test set of 284 unique patient cores, we achieved testing accuracies of 79.3% and 89.7% for 4-class (0, 1+, 2+, 3+) and 2-class (0/1+ , 2+/3+) HER2 classification, respectively. BlurryScope automates the entire workflow, from image scanning to stitching and cropping, as well as HER2 score classification.

## Introduction

The advent of digitization in the field of pathology has drastically aided the medical workflow of histological and cellular investigations[1,2]. Pathologists can now handle greater volumes of patient data with higher precision, ease and throughput. The digitization of biopsy tissue slides, for instance, has led to numerous favorable outcomes, with ameliorations in remote assessments, file transfers, research, analysis ergonomics, and overall patient care[1]. Computational advantages notwithstanding, a set of drawbacks accompany these enhancements, namely the speed, cost, and size of imaging hardware[2,3]. State-of-the-art pathology scanners have speed metrics constrained by multiple factors, such as camera frame rate, stage stability, and slide exchange processes[4]. The speed of conventional microscopes, irrespective of various efforts at acceleration, e.g., through illumination manipulation[5], line scanning[6], multifocal plane imaging[7] or time-delay integration (TDI)[8], remains stunted by mechanical and optical complexities. Despite the increasing adoption of digital pathology systems, their global availability remains limited due to high costs, infrastructure requirements, and the technical expertise needed for implementation and maintenance[9]. Price ranges for state-of-the-art digital pathology scanners average more than $100K and come in cumbersome dimensions, making them difficult to acquire in resource-limited institutions or in modest, short-staffed clinics[10]. Additionally, to ensure continuous operation and backup during mechanical shutdowns, at least two digital scanners are required for each pathology department. To address this disparity, new cost-effective and compact microscopy solutions are necessary to democratize access to advanced pathology gear[11,12].

The involvement of artificial intelligence (AI) as a tool to address some of these challenges has already shown promise in certain biomedical applications[13-21]. With the ever-increasing advancements in machine learning (ML) algorithms, numerous longstanding biomedical impasses have finally been breached[22]. An especially attractive tactic of hardware compromise following digital compensation has recently seen auspicious traction[23]. Given the current scope of reconstructive powers with deep learning models, unprecedented reworkings of essential optical components have seen various successful implementations[24-28].

Here, we present a cost-effective and compact digital scanning optical microscope, termed BlurryScope, that incorporates AI-driven image analysis. As a scanning digital microscope, BlurryScope is precisely devised to achieve rapid scans of tissue slides at a markedly reduced cost and form factor than standard commercial alternatives. To rigorously ascertain the pragmatic boundaries of BlurryScope's viability, we selected the important and ambitious target of human epidermal growth factor receptor 2 (HER2) tissue classification. Breast cancer (BC) remains one of the most common cancers globally, the most prevalent among women, and a leading cause of cancer-related deaths[29]. Accurate histological diagnostics, including determining HER2 status, are essential for effective BC management[30]. HER2 expression levels are crucial for assessing the aggressiveness of BC and guiding treatment decisions. However, traditional manual HER2 evaluation is time-consuming and prone to variability[31]. The application of BlurryScope in the automated classification of HER2 scores on immunohistochemically (IHC) stained breast tissue sections thus serves as a clinically relevant and complex task for BlurryScope, and as a potentially impactful auxiliary tool for medical practices. Therefore, as part of this proof-of-concept test, we explored automated, deep learning-based HER2 score classification using BlurryScope, which contrasts notably with the conventional pathology pipeline, involving large, costly microscopy equipment, as illustrated in Figure 1.

BlurryScope automates the entire workflow, encompassing scanning and stitching to cropping regions of interest and finally classifying the HER2 score of each tissue sample. To evaluate BlurryScope's performance, we used tissue samples from tissue microarrays (TMAs) containing >1400 cores corresponding to different patients, which were split into 1144 cores used for training/validation and 284 cores used for blind testing. Images of each tissue sample were collected using a brightfield setup equipped with a CMOS camera. The system employs a continuous linear scanning method, where the microscope stage moves at a speed of 5,000 µm/s and an area throughput of ~2.7 $mm^2/s$, while the camera captures frames at a fixed rate. This scanning process introduces bidirectional motion blur artifacts. These compromised images were then used to train our Fourier-transform-based[32] neural network models for automated HER2 score classification through blurred images. A blinded test set of 284 unique patient cores was used to assess the accuracies of 4-class (HER2 scores: 0, 1+, 2+, 3+) and 2-class (HER2 scores: 0/1+, 2+/3+) classification networks, achieving 79.3% and 89.7% classification accuracy, respectively. Each patient's specimen was scanned three times to rigorously evaluate the reliability of the BlurryScope system, which revealed that the overall HER2 score consistency across all the tested cores was 86.2%, indicating a high level of repeatability in BlurryScope's classification performance, despite random orientations of the tested slides in each run. Our results and analyses highlight BlurryScope's ability to supplement existing pathology systems, reduce the time required for diagnostic evaluations, and advance the accuracy and efficiency of cancer detection, categorization, and treatment in resource-limited settings.

## Results

### Design of BlurryScope

BlurryScope is designed to be a fast, compact, and cost-effective imaging system. The optical architecture includes components adapted from a dismantled M150 Compound Monocular AmScope brightfield microscope with an RGB CMOS camera, integrated into a custom 3D-printed framework. The system is driven by three stepper motors for precise stage movement, achieving a stable lateral scanning stage speed of 5,000 µm/s with a 10x (0.25NA) objective lens. The other main optical components include the condenser and LED illumination. The structural parts were printed using SUNLU PLA+ filament, maintaining the total component cost under $650 for low-volume manufacturing (see Table 1 and Methods section, 'BlurryScope design and assembly').

It is important to note that BlurryScope is not intended to replace traditional pathology scanners used in digital pathology systems. However, it should be considered a cost-effective alternative for routinely performing specialized inference tasks where trained neural networks can provide rapid, automated and accurate information regarding tissue specimens, such as the HER2 score classification that is the focus of this work. The key performance trade-offs of BlurryScope involve concessions in resolution, signal-to-noise ratio (SNR), and the detection of smaller objects, prioritizing speed, affordability, and compact design. While these limitations may prevent its use as a standalone diagnostic tool, BlurryScope remains valuable as a complementary system, enabling preliminary assessments, assisting in triage, and expanding accessibility in settings where conventional high-end digital pathology scanners may be unfeasible or unavailable.

To shed more light on the specifications of BlurryScope, we report several parameters, including cost, speed, weight, and size in Table 1. Traditional digital pathology scanners can perform diffraction-limited imaging of tissue specimens at extreme throughputs and form the workhorse of digital pathology systems; however, their versatility and powerful features come with significantly higher costs, with prices ranging from $70,000 to $300,000[33], making them harder to scale up, especially in resource-limited environments. BlurryScope's minimalist design (dimensions: 35 x 35 x 35 cm, weight: 2.26 kg) enables rapid imaging of samples while minimizing spatial constraints in laboratory environments.

### HER2 IHC tissue imaging

To demonstrate the efficacy of BlurryScope, a total of 10 HER2-stained TMAs were used. The training and testing datasets consisted of 1144 and 284 unique patient specimens (tissue cores), respectively. Each patient sample was scanned three times (non-consecutively) to assess the repeatability of the approach, with a total duration of 5 minutes per scan (~2.7 $mm^2/s$), which covers the entire section of the tissue microarrays of the slide, including the empty space between the cores. This extensive dataset allowed for a comprehensive evaluation of BlurryScope's capabilities in automated HER2 scoring. The standard of comparison was the output of the same set of slides imaged with a state-of-the-art digital pathology scanner (AxioScan Z1, Zeiss)[37].

The stitch generated with a standard scanner (Supplementary Figure 1a) has a clear and crisp delineation of all the cores since the scan undergoes a "stop-and-stare" operation. That is, the stage is physically halted for the duration of each camera acquisition. In contrast, Supplementary Figure 1b demonstrates BlurryScope's continuous scanning output by capturing images at a running lateral stage speed of 5,000 µm/s in a zigzag fashion (Supplementary Figure 2). This rapid acquisition introduces bidirectional motion blur artifacts. Though there is a widening and smudging of features due to the effect of motion blur, the individual cores are still fully separated in the final stitched mosaic, which allows for automated cropping and labeling of each patient tissue core.

The scanned tissue images corresponding to different HER2 scores (0, 1+, 2+, 3+) for individual patient cores are compared in Figure 2. Figure 2a shows the results from a traditional pathology scanner, yielding sharp, well-defined images for each HER2 score. In contrast, Figure 2b presents the results from BlurryScope, with images exhibiting opposing directions of blur. Despite the smearing of various details, some correspondence between both image descriptions is still discernible. Lower-scored HER2 images exhibit fewer brown hues and less geometrical heterogeneity compared with higher-scored ones. This suggests how HER2 classification tasks may still be successful on such compromised data. It is important to note that the interpretation of IHC stains is dependent on both cellular location and intensity. Most IHC stains highlight the cell nucleus and have only two levels, distinguishing between positive and negative staining. However, HER2 staining is different as it follows a four-level scoring system (0 to 3+) and is evaluated based on membrane expression rather than nuclear staining and variable levels of stain intensity. This distinction introduces greater inter- and intra-observer variability among pathologists, as the assessment depends on both *stain intensity and continuity* across the membrane. Additionally, the quantification of HER2 positivity is limited to invasive tumor cells, explicitly excluding carcinoma in situ, even when these cells exhibit a similar HER2 staining pattern. These aspects of the HER2 evaluation add further complexity to the interpretation process.

Given the critical role of stain intensity and structural integrity/continuity in HER2 assessment, the endurance of these facets within a motion blur compromise is essential for precise classification. In our previous work, GANscan[28], a conditioned generative adversarial network (GAN)-based deblurring approach, was used to reconstruct high-speed continuous-scan images of H&E-stained breast cancer tissue. The reconstructed images achieved an SSIM (structural similarity index measure) of 0.82 and a PSNR (peak signal-to-noise ratio) of 27 when compared to stop-and-stare control images, confirming our ability to restore fine tissue architecture and cellular morphology using trained neural network models. This approach successfully retrieved sub-micron structural details, including nuclear contours and tissue organization, which are critical for pathology assessments.

**Automated classification of HER2 scores using BlurryScope images**

Our data processing pipeline begins by automatically organizing the images of each patient sample into multi-scale stacks (Figure 3). The process starts with scanning the biopsy slides and recording them in video format using BlurryScope. These BlurryScope videos are then processed through automated stitching and labeling algorithms, which seamlessly integrate the frames into a whole-slide image. Subsequently, the individual cores are arranged into a concatenated stack of subsampled and randomly cropped patches, ensuring that the image data are both precise and representative. The resulting data are then processed by a classification neural network, configured for either 4-class (0,1+,2+,3+) or 2-class (0/1+ vs. 2+/3+) HER2 scoring (see Figures 3a-e). This approach allows for the efficient handling of complex image data and ensures the repeatability of the classification process (see Methods for details on 'BlurryScope image scanning, stitching, and cropping and labeling').

Upon finalizing both of the HER2-score classification networks (see Methods for implementation details), we ran our trained models on the blind test sets imaged by BlurryScope, covering N = 284 unique patient specimens/cores never seen before in the training phase. Since each slide was scanned three times, we were able to use this extra data to improve final accuracy results; see Figures 4-5. These multiple scans also enabled us to assess the consistency of HER2 classification results across repeated measurements for the same tissue core. We quantified the degree of variability that might arise from factors such as slide insertion, alignment differences, and potential fluctuations in the scanning process itself. To achieve this, we calculated the prediction consistency for each core by comparing the classification results across the three scans. Specifically, for each core, we identified the most frequently occurring prediction category (i.e., the mode) among the three scans and then determined the proportion of predictions that matched this mode. The results revealed an overall consistency of 86.2% across all scanned cores, demonstrating a high level of repeatability in BlurryScope's classification performance. As displayed in a bar graph of prediction consistency for each core (see Supplementary Figure 3), the majority of the cores exhibit strong consistency, where at least two out of three results have the same score, though some variability is present. This suggests that, while the model performs reliably for most samples, there are still certain cores where predictions are less stable, possibly due to factors like slide placement or operational conditions.

As detailed in the following analyses, three different distributions based on our triple measurements were evaluated for both HER2 classification networks: 1) total scans (3N), 2) maximum confidence interval (CI), and 3) average CI. Total scans include all the measured 3N images, while the highest CI method selects the result with the highest overall CI value from the three repeats, and the average CI method uses a CI-weighted calculation. This weighted CI calculation involves multiplying each score by its corresponding CI, summing the results, and rounding the final value (see Methods section, Sample preparation and dataset creation).

One way to heighten the reliability of our BlurryScope-based HER2 classification system is by leaving out results with low CI values and excluding them from the final assessment. To evaluate the balance between CI selection and accuracy vs. left out (indeterminate) percentages, we plotted their relationship for each data distribution and classification case. Figure 4 shows that, as expected, the accuracy is proportional to the CI threshold score chosen, and the number of patients left out as indeterminate cases. Figure 4a shows the testing accuracy and indeterminate percentages for the 4-class case with 3N samples, while Figures 4b-c present the same relationship for the highest CI and average CI, respectively. These figures illustrate how the chosen CI threshold value begins to exclude indeterminate patients starting around the 50% CI value mark. Figure 4d displays the testing accuracy and indeterminate percentages for the 2-class network with 3N samples, while Figures 4e-f present the same relationship for the highest CI and average CI.

In all these cases, there is a notable rise in the HER2 classification accuracy, along with indeterminate cases for CI selections above the 50% mark. A 5% improvement in HER2 classification accuracy in this range corresponds to ~10% increase in the number of indeterminate cases. This suggests that once the CI value exceeds 50%, the user should be mindful of pursuing further improvements in accuracy, as they may result in substantial increases in dropout rates with indeterminate results. Overall, these analyses serve to illustrate that BlurryScope can achieve a high testing accuracy with a manageable percentage of indeterminate results.

The classification accuracies for both networks (4-class and 2-class HER2 inference) were also evaluated with confusion matrices, as shown in Figure 5. We selected threshold CI values based on the plots in Figure 4 corresponding to a 15% indeterminate rate - indicated by the grey dashed lines, which were empirically selected. The confusion matrix for the 4-class HER2 score inference of all the acquired BlurryScope images (3N) has a testing accuracy of 75.3% based on a 15% indeterminate CI threshold. Confusion matrices were also generated for highest and average CI scores (Figures 5b-c), achieving HER2 score classification accuracies of 78.9% and 79.3%, respectively. Compared to automated HER2 classification results[37] using microscopic images from a standard digital pathology scanner, these numbers prove competitive in performance, lagging only by a margin of ~8-9%.

Figure 5d represents the confusion matrix of all tissue scans for the 2-class HER2 classification network, where 0 and 1+, and 2+ and 3+ groups are merged together, combining the two lowest and highest scores; these upper- and lower-bound categories are known to pathologists to have highly nuanced distinctions that are often difficult to differentiate. For this network, there is a markedly higher testing accuracy of 88.4% for a 15% indeterminate rate. When using the averaging CI method, the testing accuracy is slightly better, as shown in Figure 5f, reaching an accuracy of 88.8%, and for the highest CI method, the accuracy increases even further to 89.7%. For this model, the lower-right sections of the confusion matrices, which represent correctly identified negative cases, consistently show higher values compared to the upper-left sections, where true positive cases are recorded. This suggests the model is better at correctly identifying negative cases, reflecting higher specificity. On the other hand, the relatively lower numbers for positive cases indicate that sensitivity is slightly lower, meaning the model misses more true positives. This observation is important to note because while the model effectively avoids false positives, it could potentially overlook some true positive cases, which would be critical to capture in medical diagnostics.

The receiver operative characteristic (ROC) curves were also plotted for these 2-class cases (Supplementary Figure 4) and demonstrate varying balances between sensitivity and specificity across different methods. The area under the curve (AUC) is a key metric used to evaluate the overall performance of an inference model, with higher AUC values indicating a better ability to distinguish between classes. The maximum CI method, with an AUC of 0.76, achieves the best performance, indicating a strong capability to maximize sensitivity while minimizing false positives. The absolute average CI distribution (not CI-weighted), with an AUC of 0.74, performs similarly, slightly trailing the maximum CI approach but still maintaining a favorable balance. Overall, the maximum CI approach emerges as the most effective, achieving a decent balance between specificity and sensitivity, as reflected by its higher AUC. These analyses and results collectively indicate that BlurryScope is a promising digital imaging platform for quick inference of tissue biomarkers to potentially prioritize urgent cases or to streamline pathologists' busy workflow.

## Discussion

We demonstrated the utility of BlurryScope for automated HER2 scoring using a compact, cost-effective and rapid scanning microscope. Our results for automated HER2 scoring on TMA slides using BlurryScope are concordant with those obtained from a high-end digital pathology scanner[37], although the latter shows improved performance. The framework of BlurryScope presents a potentially scalable tool for digital pathology applications, particularly in environments with constrained resources. Traditional pathology scanners are substantially expensive, with costs ranging from $70,000 to $300,000. This prohibitive price is further compounded by the necessary procurement of multiple scanners to ensure continuous operation in clinical departments, posing a significant financial burden. In contrast, BlurryScope's total component cost is $450- $650 (accommodating for differences in material prices over time and in different economic climates), making it highly affordable for various resource-constrained settings. Furthermore, BlurryScope's compact dimensions and lightweight design enhance its practicality for use in medical settings with limited space and resources.

It is, however, important to recognize the limitations of this technology. BlurryScope can serve as a go-to inexpensive device for quickly sensing various tissue biomarkers and classifying features, but it is not designed to fully replace commercial scanning optical microscopes. The primary performance trade-offs of BlurryScope include sacrifices to resolution, SNR, and smallest detectable object size in favor of speed, cost and form factor. These compromises mean that BlurryScope may not capture the finer details necessary for certain diagnostic applications, and thus could primarily be used as a supplementary tool rather than a standalone digital pathology solution in clinical settings. To mitigate some of these limitations, AI may be further integrated to potentially handle resolution loss, poor SNR levels, and sensitivity issues inherent in BlurryScope. By leveraging advanced deep learning algorithms, we can enhance the quality of images and possibly improve classification accuracy further. Once powered by GANscan[28] or similar AI-based image reconstruction methods,[23] we believe that BlurryScope, with additional neural networks trained on motion-blurred histopathology images, can effectively recover nuclear morphology, cytoplasmic staining patterns, and sub-cellular details essential for feature-based diagnostic methods. Such a capability could extend its potential beyond HER2 quantification to applications such as nuclear pleomorphism grading, mitotic count assessments, and general tissue architecture evaluation. However, it is also important to acknowledge that AI systems can hallucinate, generating false or misleading information. This would raise the need for autonomous hallucination detection mechanisms to ensure the reliability of BlurryScope in unsupervised settings. Clinically deployed BlurryScope systems would thus need to incorporate robust validation protocols and uncertainty quantification metrics to address some of these concerns.

Moving forward, the prospects for enhancing and building on BlurryScope's capabilities are extensive. Because this is an inherently scalable technique based on exposure time and deep learning power, scanning speeds can be considerably increased using stronger motors and more intricate AI networks. Applying this technology to analyze other sample types, such as blood smears, bacterial specimens, or defects, also holds significant promise. Additionally, BlurryScope's programmable z-axis feature could be leveraged for continuous three-dimensional (3D) sample imaging and sensing. Such a vertical scanning capability could challenge existing deep learning-enabled autofocusing methods, providing faster and more accurate 3D imaging.

In general terms, the concept of trading off data quality for hardware bargains can be integrated with various components for diverse applications. These include soft optics[38], AI-optimized filter cubes, diffractive deep neural networks ($D^2NNs$)[39-46], microfluidic setups[47,48], and spatial light modulators (SLMs)[49], among others. The scanning methodology implemented in BlurryScope can thus be adapted for a range of histological imaging applications, including patient screening and biomarker quantification.

BlurryScope is an automated imaging device that may serve as a foundation for different modalities and specimen types. Input from board-certified pathologists could also aid in refining BlurryScope's processing steps for improved histopathological assessment accuracy. BlurryScope's most immediate application would likely be in triaging and identifying questionable or tricky sample areas. In this capacity, BlurryScope is poised to excel in the short-term future, potentially providing critical support in clinical diagnostics.

## Methods

### BlurryScope design and assembly

BlurryScope design process utilized Autodesk Fusion 360 for creating the detailed 3D models of the BlurryScope components. These models were then used to print the necessary parts on a Creality Ender 3 Pro 3D printer, ensuring precision and durability. The basic optics of BlurryScope were adapted from a M150 Compound Monocular AmScope brightfield microscope with an RGB CMOS camera (30fps at 640x480 resolution, pixel size of 5.6 µm, and a sensor size of 3.59 x 2.69 mm). Also included is a condenser, 10x objective (0.25NA) and LED illumination (Figure 1b; Supplementary Figures 5a-b). The condenser-lens system is aligned to provide uniform illumination across the sample, reducing artifacts and ensuring consistent light distribution across the field of view. The LED driver circuitry has been modified to accommodate a 24V DC input, which is regulated down to 5V using an LM2596 DC-DC converter. This power configuration provides stable and adjustable lighting conditions, optimizing image contrast and signal-to-noise ratio for histological imaging. To ensure accurate imaging, a microscope stage calibration slide was used to align the optics and test the device. For outfitting mechanical components, micro switches were used for the end stops of the linear actuators and two lead screws and couplings were incorporated for precise movement control. The system also included three stepper motors to drive the linear actuators. The X and Y axes are controlled by synchronized stepper motors mounted onto a 3D-printed base plate. The objective holder is mounted on a motorized Z-axis that provides fine-focus adjustments, enabling controlled movement of the optical configuration relative to the sample. This system minimizes mechanical backlash while enhancing positional accuracy, ensuring that scanned tissue sections remain precisely aligned throughout the imaging process. The structural parts were printed using SUNLU PLA+ filament. Altogether, the optical components and surrounding materials amount to < 650 USD. Supplementary Figure 5 shows how the various 3D-printed components fit together into the whole system along with the corresponding optics.

The integration of these components allowed for the creation of a custom-designed framework that maintained the integrity of the imaging process and enabled programmable stage and objective movement. The device is powered and controlled using Thonny (v4.14, Aivar Annamaa)[50], a Python-integrated development environment (IDE), simplifying the deployment of the control software for the stage and imaging system. The stage was programmed to move in a zigzag configuration (not raster or row-major – see Supplementary Figure 2) at 5,000 µm/s with at most 20% frame overlap using a 10x (0.25NA) objective. Although a higher NA objective and faster speeds could have been implemented, conservative stability and processing measures were taken to avoid potential mechanical failures in the long run.

The compact nature of the design allows for various additions, including the potential integration of automated objective switching, improved autofocus mechanisms, and higher numerical aperture objectives to expand imaging capabilities. Additionally, the system's flexible framework could allow researchers to implement custom modifications tailored to specific imaging tasks, such as high-speed fluorescence imaging.

**BlurryScope scanning**

The scanning and stitching process in BlurryScope is fully automated using Thonny and MATLAB (vR2022b, MathWorks, Inc), respectively. The Thonny program controls the motorized stage to ensure precise movement and continuous image acquisition. The AmScope camera software is used simultaneously with this to record videos of the samples during the scanning process. The software is not synced or coordinated mechanically with the stage. The scan follows a zigzag geometry and generates rows of opposing blur width. A motion-blurred image is a function of the stage speed and the camera acquisition time[28], as follows,

$$(1) \qquad \underline{I}(x, y; t) = I(x + s_x t \pm \delta_x(t), y \pm \delta_y(t))$$

where $I(x, y)$, at stage speed $s_x$=0, is an image at rest. The x-translated blurry image, $\underline{I}$, has the following time dependence[28]:

$$(2) \qquad \underline{I}(x, y) = I(\frac{x}{s_x}, y \pm \delta_y) \textcircled{V}_{\frac{x}{s_x}} \prod(\frac{x \pm \delta_x}{s_x T})$$

where $T$ is the camera acquisition time, $\textcircled{V}_{\frac{x}{s_x}}$ represents the convolution operator over the variable $x/s_x$, which has dimensions of time, and $\prod$ is the 1D rectangular function with a width of the blur distance $s_x T$. $\delta_x(t)$ and $\delta_y(t)$ model the system's ringing-induced jitter, which can be represented as small and oscillatory deviations along the x or y axes. Eq. (2) encapsulates the physical effect of spatial smearing as the result of a convolution operation. Thus, the blurred image is, in essence, its counterpart crisp image convolved along the direction of the scan by a rectangular function, with a width dictated by to the acquisition time. For a scanning speed of $s_x = 5{,}000 \frac{\mu m}{s}$and $T = 7.8ms$,we have $s_x T = 39\mu m$. This means that the continuous overlap of consecutive frames covers a margin of $39\mu m$, in opposite directions for each row of scanning. The exposure time here was selected based on the 1W brightness capacity of the LED and the blur that our models could reliably tolerate at the desired scanning speed. It is important to note that a pulsed LED could have been implemented here by synchronizing the frequency with the camera's frame rate and the stage motion. This coordination would ensure complete tissue coverage without spatial gaps and allow for a reduced motion blur while maintaining scanning continuity. The total scanning duration for a whole TMA slide is under 5 minutes, with a throughput of ~2.7 $mm^2$/s. This figure is calculated based on the fact that each captured frame covers a field of view (FOV) of 640 μm × 480 μm, so the total area of a single frame is 307,200 $\mu m^2$. Since the stage movement dictates the rate at which new regions are imaged, the effective area scanned per second is determined by the stage velocity along the horizontal direction of motion. Thus, the scanned area per second is given by (5000 μm + 640 μm) × 480 μm, which yields ~2.7 $mm^2$/s.

## BlurryScope image stitching

The stitching of the captured video frames into whole-slide images is achieved using an automated algorithm that combines correlation and square wave-fitting methods. This approach ensures that the images are accurately aligned and stitched together, even when there are slight variations (jerks) in the movement of the stage. The stitching algorithm operates by analyzing the correlation between consecutive frames to classify each frame in the video file as either "moving" or "static." It identifies windows of frames where the average correlation is very high, indicating a stable scan line. In our optimized zigzag scan pattern, there is a 0.5-second pause before and after each scan line (~30 frames) and a vertical jump to the next scan line (~3 frames). These pauses are detectable through correlation analysis. However, the algorithm can encounter errors in regions of the slide without distinct features, where motion cannot be inferred from the video alone. To address this, the algorithm incorporates a model of the zigzag motion pattern. This model labels each frame as "moving" or "not moving," fitting a square wave to the data as the motion alternates between scanning (moving) and pauses (not moving) at the start and end of each scan line. Using this refined data, the algorithm can accurately identify the start and end of each scan line, enabling the generation of a stitch from the video frames (Supplementary Figure 6). The BlurryScope stitching operation is carried out entirely in MATLAB and takes ~2 minutes for a single slide (7.8 $mm^2/s$) on a GeForce 4090 RTX graphics processing unit.

## Tissue image cropping and labeling

Our automated algorithm to crop and label tissue cores utilizes a spreadsheet that contains HER2 scores for each tissue sample on our TMAs that were independently verified by three certified pathologists. This is our source for ground truth labels. The process begins with a color correction (white balance) of each stitched TMA image. This is achieved by sampling a region devoid of tissue and other artifacts and subtracting it from the image, ensuring a uniform white background across all data. A grayscale copy of this image is then created, thresholded and blurred. The border is also cleared to facilitate the accurate drawing of contour boxes over tissue samples. The outermost coordinates of these contour boxes are saved, yielding a refined rectangular region that neatly encapsulates the tissue samples. The selected region is then segmented based on the row and column coordinates provided by the layout diagram accompanying the slides from TissueArray. This diagram includes detailed metadata specifying the position and patient characteristics of each tissue core. Each grid cell is aligned with its corresponding tissue core, ensuring accurate labeling and matching for further analysis. The isolated tissue core image within each box is then saved and assigned a label ranging from 0 to 3+, which is derived from the pathologist-verified spreadsheet. Upon completion of the labeling process, taking 10s per slide (94$mm^2/s$), we transition into the dataset creation phase.

## Sample preparation and dataset creation

Histological samples were acquired from TissueArray, specifically breast tissue sections, and were stained for HER2 using standard IHC staining protocols at the UCLA Translational Pathology Core Laboratory[37]. The array set included 10 TMA slides with the following catalog numbers: BR1141a, BR1202, BR1503f, BR2082c, BR1507, BR10010f, BR20812, BR20814, BR20815, and BR20826, allowing for structural reference and reproducibility. These samples were then scanned using BlurryScope in sets of 3 repeats with different orientations and setup orders to account for various sources of variability. The dataset creation involved capturing multiple video frames of each sample, which were then processed and stitched together. This dataset was then used to train and validate the HER2 score classification networks. Each training instance was a combination (3D concatenation) of 1 fully downsampled 512x512 image and four randomly cropped sections of the same 512x512 dimensions from the original stitched result (a 5125x512x3x5). The full dataset consisted of 3-fold scans of 1144 unique patient specimens (cores) for training and 284 for testing – i.e., 3N = 4284 core images in total with all the repeats.

Three distinct distributions were evaluated for both networks: 1) total scans, 2) maximum CI, and 3) average CI. The total scans distribution incorporates all measured images across the three repeats, providing a comprehensive overview of all available data. The maximum CI method selects the result from the repeat that has the highest overall CI value, ensuring that the most confident prediction is used for that image. In contrast, the average CI method takes a more nuanced approach by calculating a CI-weighted average. This involves multiplying each score by its corresponding CI value across the three repeats, summing these products, and then rounding the final result to provide a balanced prediction that accounts for all measurements while weighing them according to their corresponding confidence. This approach ensures that the final score reflects not just the raw predictions but the reliability of each repeat.

**HER2 Score 4-Class and 2-class classification network architectures and training scheme**

The classification network for 4-class HER2 scoring was based on an eFIN architecture[32,51], a Fourier-transform-based network that manipulates the spatial frequency domain information using dynamical linear maps. In adapting the eFIN architecture for this application, we introduced two critical modifications: (1) a global average pooling (GAP) layer was appended to the final convolutional block, which reduces spatial dimensions to a 4-element vector aligned with our four-class taxonomy; and (2) a cross-entropy loss was adopted to optimize probabilistic alignment between our predictions and the ground truth. These targeted adjustments maintain eFIN's core feature strengths while optimizing task-specific classification through dimensionally matched outputs and statistically grounded optimization. The classification network for 2-class HER2 scoring was trained with hyperparameters and variables that were similar to the 4-class case, following the original eFIN architecture[32]. The classification networks' training was optimized using an AdamW optimizer with a weight decay factor of $10^{-4}$. The training commenced with an initial learning rate set at $10^{-3}$, which was dynamically adjusted using a cosine annealing scheduler with warm restarts. The training and testing operations were conducted on a desktop computer equipped with a GeForce RTX 3090 graphics processing unit, 64GB of random-access memory, and 13th Gen Intel Core™ i7 processing unit. The classification networks were implemented using

PyTorch, with a single testing core image stack taking about 0.85 seconds to classify.

## Code Availability

The testing codes, the testing model, and testing example images can be found at: https://github.com/hlc1209/BlurryScope

**Correspondence and requests for materials** should be addressed to Aydogan Ozcan, ozcan@ucla.edu

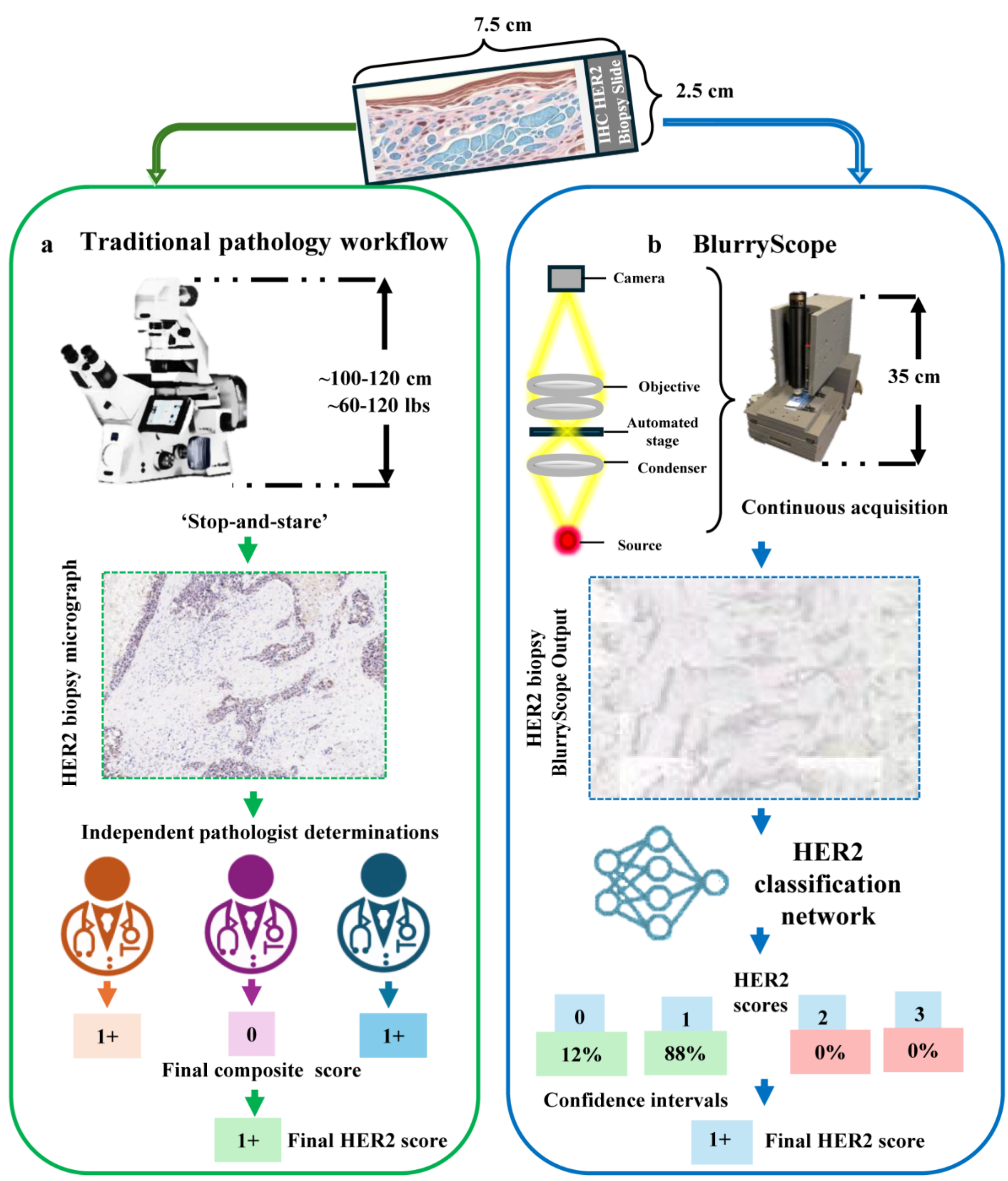

**Figure 1: Comparison between traditional digital pathology workflow and BlurryScope for automated HER2 classification.** **a**, Traditional pathology workflow involves inspection and analysis of tissue sections using a digital pathology scanner. **b**, BlurryScope automates the entire workflow from image acquisition to HER2 classification by integrating specialized optical hardware and deep learning algorithms to rapidly process motion-blurred images with deep learning-based HER2 classification, offering an efficient and cost-effective alternative for task-specific inference in resource limited settings.

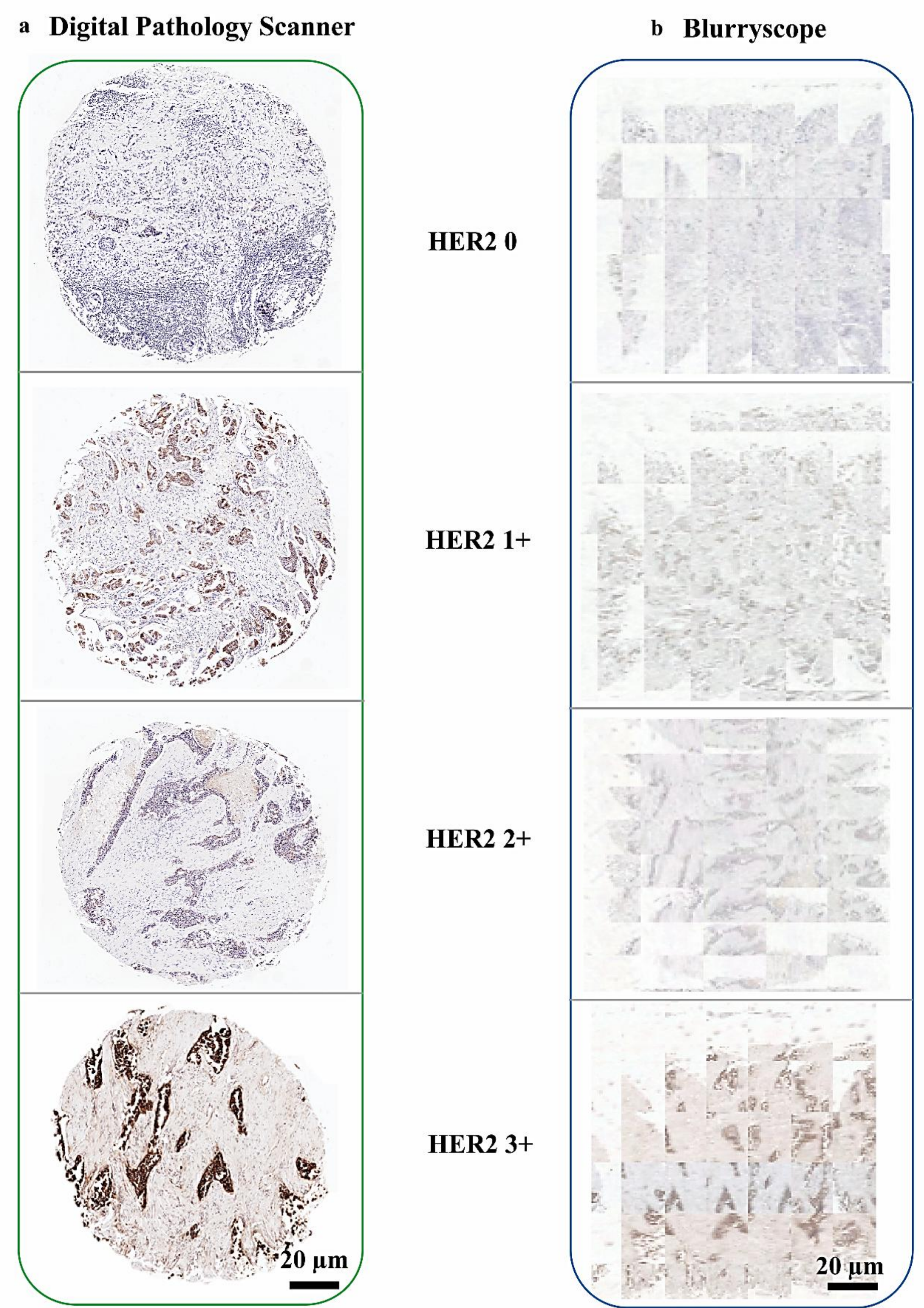

**Figure 2: Images of tissue cores with different HER2 scores. a** Images of tissue specimens with HER2 scores (0, 1+, 2+, 3+) obtained using a traditional digital pathology scanner, showing clear and well-defined cores for each HER2 score. **b**, Same as **a**, except the images are obtained using BlurryScope, where the images exhibit smudged details.

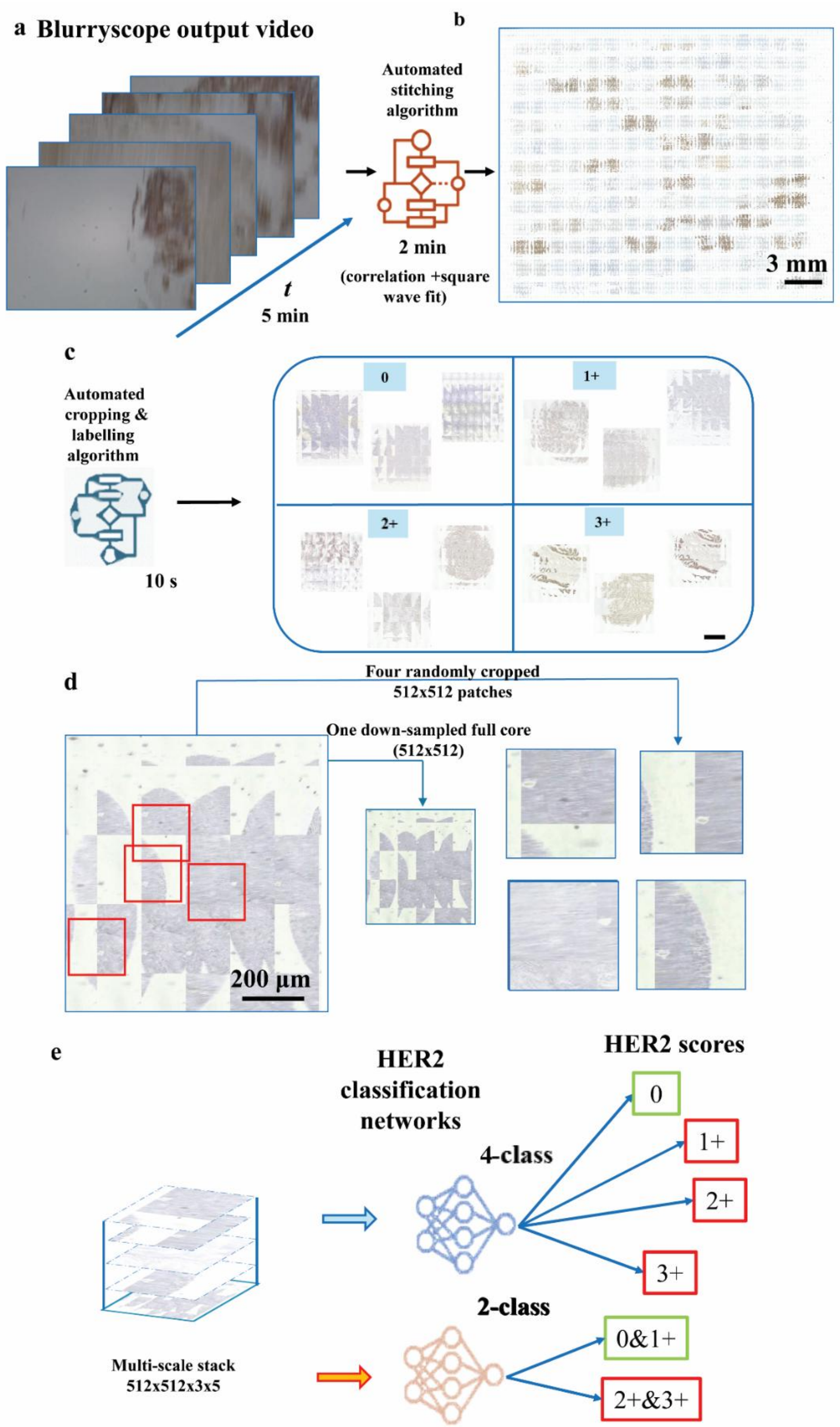

**Figure 3: BlurryScope data processing pipeline. a**, The data processing workflow of BlurryScope begins with the continuous video output of the scanned slides, followed by **b**, automated stitching and **c** labeling. **d**, Images are then cropped and concatenated into a stack of subsampled patches. **e**, These image patches are then processed by deep learning-based classification networks. Scale bar 200 μm.

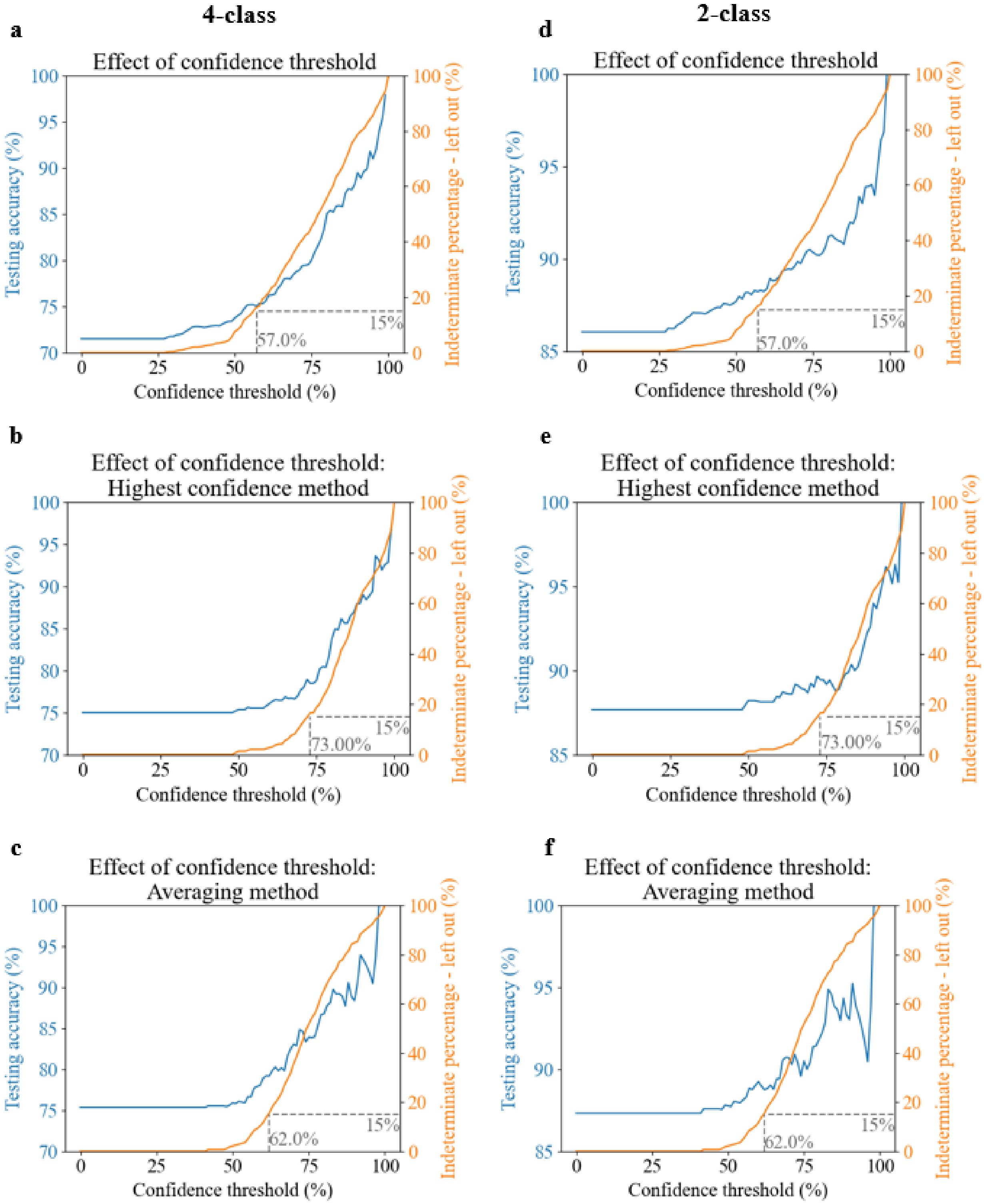

**Figure 4: Testing accuracy as a function of the confidence threshold. a**, Testing accuracy and indeterminate percentage for the 4-class HER2 classification system with 3N samples. **b**, Testing accuracy and indeterminate percentage for the 4-class HER2 classification system with the highest CI. **c**, Testing accuracy and indeterminate percentage for the 4-class system with the CI-

weighted method. **d**, Testing accuracy vs. indeterminate percentage for the 2-class system with 3N samples. **e**, Testing accuracy and indeterminate percentage for the 2-class system with the highest CI. **f**, Testing accuracy and indeterminate percentage for the 2-class system with the CI-weighted method. Grey dashed lines refer to a 15% indeterminate rate.

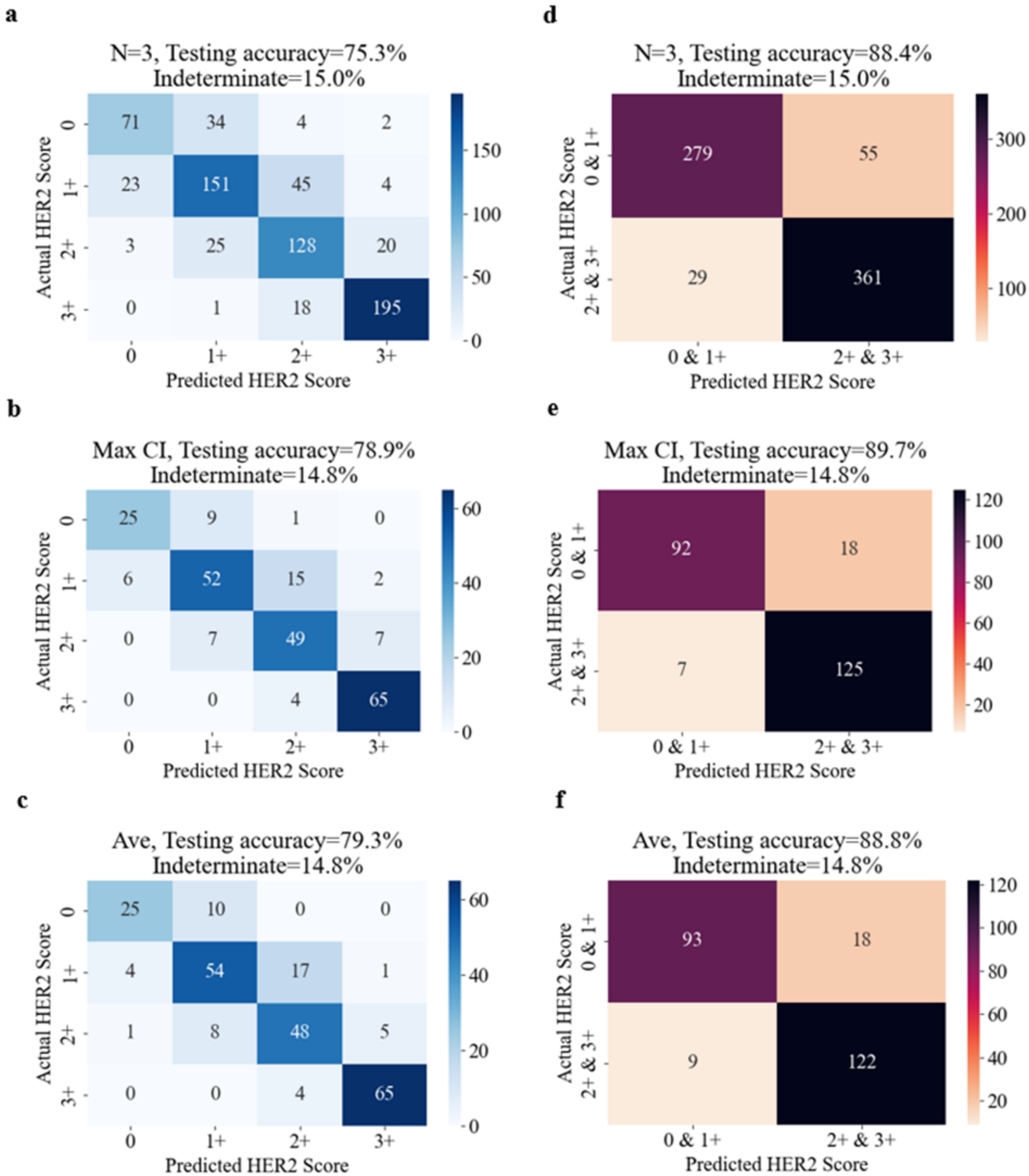

**Figure 5**: **Confusion matrices and classification accuracy**. **a**, Confusion matrix for the 4-class HER2 classification network for all scans. **b**, Confusion matrix for the 4-class HER2 classification network with the highest CI scores. **c**, Confusion matrix for the 4-class HER2 classification network with average CI scores. **d**, Confusion matrix for the 2-class HER2 classification network for all scans. **e**, Confusion matrix for the 2-class HER2 classification network with the highest CI scores. **f**, Confusion matrix for the 2-class network with the average CI scores. N refers to the number of times each slide (or patient sample) was non-consecutively and separately scanned.

**Table 1. Specifications of BlurryScope by price (USD), speed, weight and size.**

| Scanner | Price | Speed | Weight | Dimensions | Common applications | Resolution |
| --- | --- | --- | --- | --- | --- | --- |
| Digital pathology scanners[33-36] | \$70,000 - \$300,000 | 1-7 $mm^2/s$ | 34-55 kg | 52 x 52 x 62 cm – 120 x 85 x 100 cm | High-throughput whole-slide imaging, telepathology | 20-40x/0.75-0.95 NA, no blur |
| BlurryScope | \$450-\$650 | 2.7 $mm^2/s$ | 2.26 kg | 35 x 35 x 35 cm | AI-assisted pathology, motion-blur correction, rapid screening | 39 µm blur width, 10x/0.25 NA |